\documentstyle[twocolumn,aps,eqsecnum,aps,epsfig]{revtex}

\begin{document}
\draft
\preprint{CARRACK/2000-03}
\title{
Coherent time evolution of highly excited Rydberg states in pulsed 
electric field: Opening a stringent way to selectively field-ionize the 
highly excited states      
}
\author{M. Tada, Y. Kishimoto, I. Ogawa\cite{byline}, H. Funahashi$^1$,
K. Yamamoto$^2$, and S. Matsuki}
\address{
Nuclear Science Division, Institute for Chemical Research, Kyoto
University, Gokasho, Uji, Kyoto 611-0011, Japan\\
$^1$ Physics Department, Kyoto University, Kyoto 
606-8503, Japan\\
$^2$ Department of Nuclear Engineering, Kyoto 
University, Kyoto 606-8501, Japan
}
\date{\today}
\maketitle
\begin{abstract}
Coherent time evolution of highly excited Rydberg states in Rb 
(98 $\leq$ $n$ $\leq$ 150)  under pulsed 
electric field in high slew-rate regime was investigated with the field 
ionization detection.  The electric field necessary to  
ionize the Rydberg states was found to take  
discrete values successively depending on the slew rate.  Specifically 
the slew-rate dependence of the ionization field varies with the quantum 
defect value of 
the states, i.e. with the energy position of the states relative to the 
adjacent manifold.  This discrete transitional behavior of the ionization 
field observed for the first time is considered to be a manifestation of 
the strong coherence effect in the  
time evolution of the Rydberg states in pulsed electric field and opens 
a new effective way to 
stringently select a low-$\ell$ state from the nearby states by field 
ionization. 
\end{abstract}


\narrowtext

%

Highly excited Rydberg states~\cite{Gallagher} in the ramped 
electric field is one of the most interesting systems which provide 
ideal and versatile   
situations for investigating the coherence effects in the time evolution of 
a quantum system with many 
potential-energy curves crossing one another\cite{Harmin1}.  In spite of 
this interesting feature
and also of the potential applicability to the wide area of 
fundamental physics including cavity QED and quantum 
computation~\cite{Haroche}, the Rydberg 
states with high principal quantum 
number $n \gtrsim$ 80 have not been investigated in detail, partly 
because of the difficulty 
in selectively detecting a particular state from  many 
close-lying states; in such highly excited states, the field ionization 
process generally occurs both 
through the non-adiabatic and adiabatic transitions, resulting multiple 
ionization thresholds and the selective detection of a particular state becomes 
increasingly more difficult.     

The purpose of this Letter is to present the experimental 
results on the field ionization of the Rydberg states with $n = 98 
- 150$. It was observed for the first time that in high  
slew rate regime in the applied 
pulsed electric field, the field ionization process  
has single threshold value: Specifically the ionization 
electric field 
takes discrete values successively with increasing slew 
rate, and this dependence varies with 
the position of 
the states relative to the adjacent manifold. 
This transitional 
behavior in the field ionization shows regular dependence on the principal 
quantum number $n$, thus indicating that this behavior is quite general and 
applicable to a wide range of higher-lying Rydberg states. 
Since the differences in the field ionization values were found to be 
large enough, i.e. 300 \% for the 111$p_{3/2}$ and 111$s_{1/2}$ 
states, it is possible to stringently select a low-$\ell$ state from the 
close-lying states by field ionization. From these 
characteristic behaviors,  it is strongly suggested that the coherence 
in the time evolution 
under the pulsed electric field plays decisive role to the 
behavior of the field ionization. 
%
%
%
%
%
%

\begin{figure}
\epsfig{file=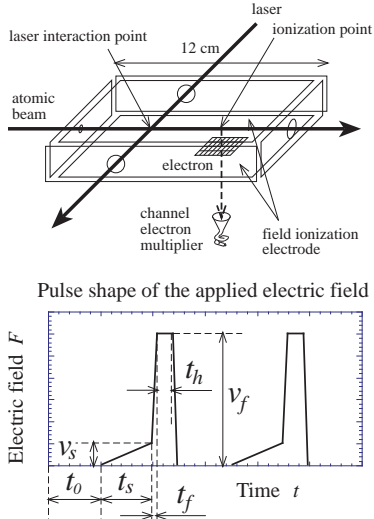,height=7cm}
\caption{Experimental setup for investigating the time evolution and the 
field ionization process under the pulsed electric field in high 
slew rate regime. Also shown is the pulse shape of the applied 
electric field.
}
\label{fig:setup}
\end{figure}

The experimental setup is shown in Fig.~\ref{fig:setup}.  Thermal 
Rb atoms in the ground-state atomic beam are passed through 
a laser excitation region and then the field 
ionization region, which are about 40 mm apart each other.  The whole 
volume of the excitation and the field ionization regions is 
surrounded with three pairs of  planer copper electrodes 
to compensate the stray field in three axes and 
also to apply the pulsed electric field for the ionization. The selective 
field ionization (sfi) electrodes consist of 
two parallel plates of 120 mm length, in one of which a fine 
copper-mesh grid was incorporated into the area of 20 $\times$ 20  
mm$^2$, thus allowing to pass and detect 
the field ionized electrons with a channel electron multiplier.  

The sfi electrodes and the laser interaction region were attached to  
a cold finger in a  
cryostat, thus the temperature can be varied from room temperature down to 
lower temperature with liquid N$_2$ and He. 

Two-step cw-laser excitation was adopted to 
excite the Rydberg $nj$ states from the 5$s_{1/2}$ ground state of $^{85}$Rb 
through the 5$p_{3/2}$ second excited state.  A diode laser (780 nm for the 
first step) and a dye laser of 
coumarin 102 excited by a Kr ion-laser (479 nm for the second 
step) were used.  The main reason 
to use the cw lasers to excite the Rydberg states 
instead of the usually adopted pulsed lasers is that 
we want to use this field ionization scheme for the selective 
ionization of highly excited Rydberg states in a ${\it continuous}$ mode as 
discussed later.

The pulse shape applied for the field ionization is shown also in 
Fig.~\ref{fig:setup}. The pulse sequence was produced with 
a waveform generator 
NI5411 and the field ionization signals were detected and analyzed 
with the LabVIEW data acquisition system on a pc computer. 
Repetition rate of the pulse was kept to 5 kHz so that the detection 
efficiency of the Rydberg states is optimum for the atoms with  
velocity of 350 m/s~\cite{eff}.  The ionization mainly occurs at 
the steep 
rise of the pulse during the time $t_f$, but the peak field can be  
kept for a time $t_h$ (holding time) to ionize also the states with 
longer life time than $t_f$ under the electric field.  This point will 
be discussed later.   

%
%
%

During the course of the present experiment, stray 
field of $\sim$ 80 mV/cm was found to appear in the interaction 
region. By applying suitable compensation potentials in three 
axes,  the stray field was reduced 
to less than 10 mV/cm. In this stray field, the $p$ state with
$n$ less than 120 is well separated from the adjacent manifold 
levels. The effect of the mixing of the low-$\ell$ states to the 
neighboring levels on the field ionization behavior is 
discussed later.   

\begin{figure}
\epsfig{file=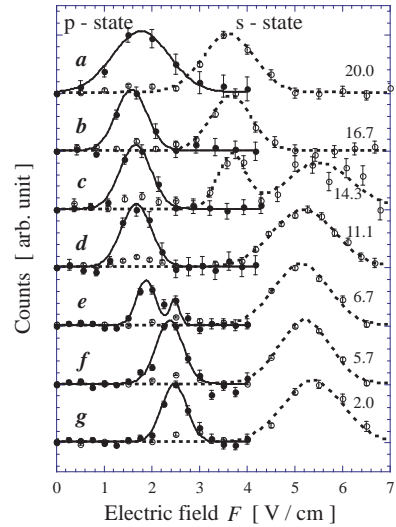,height=7cm}
\caption{Typical field ionization spectra for the 111$s_{1/2}$ and 
111$p_{3/2}$ states measured by varying the slew rate, values of which 
are shown at the right side of each spectra.  The spectra 
for the repspective  states ($s$ and $p$) taken separately were superposed upon 
each other in these spectra. Solid and dashed lines are the fitted results 
with a Gaussian plus linear background. 
}
\label{fig:sfi-spec}
\end{figure}

In Fig.~\ref{fig:sfi-spec} shown are field ionization spectra of 
111$s_{1/2}$ and 111$p_{3/2}$ states~\cite{com1} as a 
function of the applied electric field $F = v_{f}/l$ ($l$ is 
the distance of the sfi electrode), which were measured by 
varying the slew rate $S = F / t_{f}$.  Here the slow 
component of the pulsed field $v_s$ was set to zero.  

In these spectra, 
only one prominent peak was found as the threshold electric field.  
Remarkably the sfi field value at the peak changes 
to a smaller value with increasing slew rate at a particular slew rate. 
Moreover the sfi fields for the $s$ and $p$ states 
are quite different from each other at the same slew rate.   For 
example, the electric 
field necessary to ionize the 111$p_{3/2}$ state 
at the slew rate of 11 V/(cm$\cdot\mu$s) is 1.7 V/cm, while the 
value is 5.2 V/cm for the 111$s_{1/2}$ state, more than 300 \% 
difference (see the spectrum $d$ in Fig.\ref{fig:sfi-spec}).  

\begin{figure}
\epsfig{file=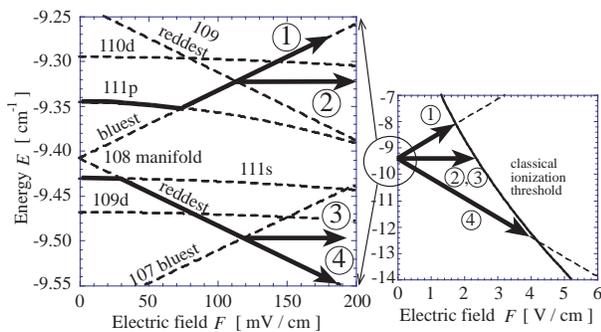,height=4.5cm}
\caption{Stark energy diagram near the 108 manifold in Rb together with the 
classical field-ionization threshold-line ($E=-6.12\sqrt{F}$), where 
all the avoided crossings are not explicitely shown for simplicity.  
The arrows 1 to 4 
indicate the possible trajectories for the adiabatic (2 and 3) and the 
extreme non-adiabatic (1 and 4) transitions.
}
\label{fig:starkmap}
\end{figure}

It should also be noted here that the 
transitional behavior in the 109$d$ state (not shown in 
Fig.~\ref{fig:sfi-spec} 
to avoid complexity) is  the same as in 
the $s_{1/2}$ state.  This means that the transitional behavior depends on 
the position of the states relative to the adjacent manifold: As seen 
from the relevant Stark energy-field diagramin in Fig.~\ref{fig:starkmap}, the 
${\it upper-positioned}$~\cite{com2} $p$ state shows 
different transitional 
behavior from the ${\it lower-positioned}$ $s$ and $d$ states.      

\begin{figure}
\epsfig{file=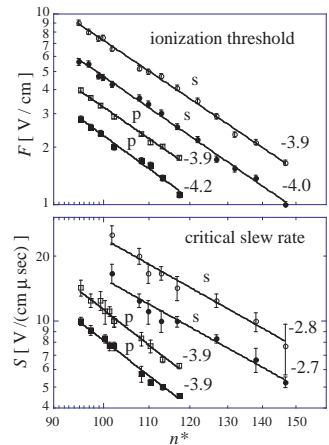,height=6cm}
\caption{Dependence of the selective field ionization (sfi) value and the 
critical slew rate on the effective principal-quantum-number $n^{*}$ in Rb 
Rydberg 
states. Discrete two values of the sfi field observed and their corresponding 
slew rates are plotted together with the fitted lines of n-dependence. 
The number to each plot is the coefficient $\alpha$ in the fitting 
of $(n^{*})^{-\alpha}$ dependence with the estimated error of $\pm 0.2$. 
}
\label{fig:n-dep}
\end{figure}

In Fig.\ref{fig:n-dep} shown is the effective 
principal-quantum-number $n^{*}$ dependence of 
the critical slew rate $S_c$ and the ionization electric fields $F_c$ for 
the $s$ and $p$ states.  Here the critical slew rate $S_c$ is defined 
as the value at which the transition of the sfi field just starts to a new 
value.    Also the sfi field  
corresponds to the peak position of the ionization signal.   
These values vary quite regularly with $n^{*}$ 
ranging from 95 to 147; approximately $ F_c \propto (n^{*})^{-4.0} $
for both the $ p $ and $ s $ states,
while $ S_c \propto (n^{*})^{-4.0} $ for the $ p $ state
and $ S_c \propto (n^{*})^{-2.8} $ for the $ s $ state.

These results indicate that this transitional behavior is quite general 
for a wide range of $n$ and 
thus can be applicable to 
selectively ionize the $s$ and $p$ states for a wide range of higher 
excited states. The difference in the ionization field is 
$\sim$~300 \% for the $s$ and $p$ states, quite large compared to the case of 
adiabatic transition in which the difference would be only 5 \%  
for the states at $n \sim 110$.   

%

Related to the selectivity in the Rydberg states, it should be noted 
that there observed some sfi signals at the lower field 
region (1.0 $\sim$ 2.4 V/cm) in the sfi spectrum of $s$ state: The 
signal peak at this portion changed discretely with slew rate as in the 
same manner as of the $p$ state.    This 
part of signal counts is due to the effect of blackbody radiations 
which induce the transition from the initial $s$ state to the $p$ state. 
In fact the counts at 
this portion measured by varying the temperature of the excitation-detection 
region from 120 K to 40 K was found to  depend linearly on the  
temperature as expected.  The observed transition rate to 
the $p$ state is in 
roughly agreement in its absolute values and in good agreement in its 
temperature dependence with the theoretical predictions.  This 
agreement indicates also that the selectivity of the excited states 
with the field ionization method in the pulsed electric field regime is 
quite good even at such highly excited region. The detailed 
discussion on the $s$ to $p$ transitions and the effect of blackbody 
radiations will be reported elsewhere. 

The above results were all obtained without the slow component 
$v_s$ of the pulsed field.  Switching on this 
value, the transitional behavior of the $s$ state remains the same, 
while that of the $p$ state changes drastically as in the 
following: When the applied slow 
component field with its slew rate less than 1 mV/(cm$\cdot \mu$s) exceeds 
the first anti-crossing field ($\sim$ 75 
mV/cm for the 111$p_{3/2}$ state, see Fig.~\ref{fig:starkmap}), the transitional 
behavior changes abruptly, 
becoming the same as of the 
$s$ state.  This means that once the first anti-crossing is traversed 
adiabatically and the state is mixed with the adjacent manifold 
levels, then the following sfi behavior for the $p$ state under the 
high slew rate regime changes completely, resulting no difference 
in their behavior between the states of opposite positions to the 
adjacent manifold.     

These experimental results, especially the transitional behavior, are 
in general not in agreement with simple predictions from the incoherent
contributions of adiabatic and non-adiabatic transition processes:   
The experimental sfi spectra have only one prominent peak and this 
peak field  does neither correspond exactly to the expected position 
from the purely adiabatic (paths 2 or 3 in Fig.~\ref{fig:starkmap}), nor 
diabatic transitions (paths 1 or 4) leading to the reddest or bluest 
trajectory in the adjacent manifold; the expected fields for these 
paths (1 to 4) in $n$ = 108 (corresponding to the adjacent manifold 
of 111$s$ and 111$p$ states from the quantum defect values in Rb) are 
1.8, 2.4, 2.4, and 4.2 V/cm respectively, 
which should be compared to the observed field values of 1.7 and 5.2 V/cm 
for the $p$ and $s$ states, respectively.       
More importantly, these spectra show very clear 
slew rate dependence as described above which can not be explained from a 
simple incoherent process.     

%

Recently Harmin~\cite{Harmin1} examined coherent time evolution on a grid of 
Landau-Zener anti-crossing under a linear ramped electric field in which 
each manifold levels are treated as linear in time, parallel, and 
equally spaced and infinite number. The time development of  an 
initially populated state is then governed by two level Landau-Zener 
(LZ) 
transitions at avoided crossings and adiabatic evolution between 
them. The key 
parameters in this model study are 1)the LZ transition probabilities $D$ for
making a non-adiabatic transition process at the anti-crossing traversals and 2) 
the dynamical phase unit $\varphi$. 
The phase unit $\varphi$ is the area covered by the pair of the 
adjacent up- and down-going levels~\cite{Harmin1}.  The overall difference 
in phase advance $\Delta \Phi$ for two paths from zero to the ionization 
field is equal to the sum of the phase units 
$\varphi$ in the whole area covered by the two paths. In the 
actual Stark energy-field grid system, the phase units $\varphi_i$ are not 
a constant but vary with respective anti-crossings.       
These parameters $D$ and $\varphi$ are estimated approximately by  
$D = [{\rm exp}(-\pi \tilde{\mu}^{2} \varphi)]^{2},  \varphi \sim (3 \dot{F} 
n^{10})^{-1}$,  where $\tilde{\mu}$ is an average low-$\ell$ quantum defect and 
$\dot{F}$ is the slew rate of the applied electric field $F$. 
          
The probability 
$D$ increases monotonically with increasing slew rate, while the 
overall difference in phase advance $\Delta \Phi$ of the state wave function through the 
grid is effective in modulo 2$\pi$ and strongly affects the coherent 
nature of the process through the 
interference between many states populated along the way of 
traversals in the applied electric field~\cite{Harmin1}.  

In the present experimental setup with slew rate $\sim$ 20 
V/(cm$\cdot\mu$s) at $n \sim$ 100, the probability 
of non-adiabatic transition $D$ is quite high, reaching $\sim 
99.8 \%$ from the above estimation so 
that our case corresponds to the non-adiabatic limit in Harmin's 
treatment in a good approximation.  

%

Taking into account the 
quantum defect values of low $\ell$ states, estimated phase unit  
$\varphi$ varies from $10^{-2}$ to 
$10^{-3}$ with increasing electric field. The number of avoided 
crossings traversed between zero field 
and the field ionization region in Rb is estimated to be 
$N_{blue} \sim 900 \left( \frac{n}{100} \right)^2,~  
N_{red} \sim 1150 \left( \frac{n}{100} \right)^2$
for the up- (bluest) and down- (reddest) going trajectories, 
respectively. At the region of $n \sim 110$, this number of 
anti-crossing traversals suggests that the phase 
advance is over $2\pi$ at some field value.   When this 
resonance condition is fulfilled, the constructive interference 
between the many number of Stark states along the advance of the 
pulsed electric field may result in one prominent peak in the sfi 
field.  In Harmin's analysis~\cite{Harmin1}, the population of the states 
under the 
electric field was found to make a series of resonances in a form of 
$\it{lanes}$ along the up- or down-going directions of the initially 
excited state at zero field.  

This 
general feature in the non-adiabatic limit seems to be satisfied 
experimentally, 
since a discrete sequence of sfi threshold field  was found, 
depending on the slew rate. However there is a significant difference 
in the experimental observations compared to the model calculation in that 
the observed sfi field values for both of the $s$ and $p$ states 
decrease with increasing slew rate. Contrary to this observation, the 
model calculation predicts that with increasing slew rate, the 
most populated state approaches to the limiting trajectory in the 
manifold, i.e. to the bluest state for the up-going initial 
state or the reddest state for the 
down-going initial state, depending on the direction (up- 
or down-going) of the initial state.  This suggests thus that the sfi 
threshold field 
for the $s$ and $d$ states 
in Rb should increase with increasing slew rate,
in disagreement with the experimental results. 

Finally we note that the sfi spectra observed with the holding time 
$t_h$ extended up to 500 $\mu$s showed no distinguishable difference from 
those taken with the short holding time of 1$\mu$s, indicating all 
the states ionized have decay lifetime shorter than 1 $\mu$s in the 
electric field.  Also we observed 
no significant sfi-signals over 10 V/cm in the spectra measured. 
The decay rate of the blue states 
in the electric field by the tunneling process, estimated from the 
hydrogenic approximation, is much higher 
than 2$\times 10^{3}$ s$^{-1}$.  Therefore the ionization 
process for the states along the blue lines is not due to the 
tunneling process but of autoionization-like 
one due to their mixing to the red continuum, even though such coupling 
becoming weaker as $n$ increases~\cite{Gallagher}.       

%

In conclusion, we observed for the first time a discrete transition of 
the threshold sfi field with 
slew rate in the highly excited Rydberg states in Rb, the behavior of 
which depends also on the position of the low $\ell$ states relative to 
the adjacent manifold.  The experimental results strongly suggest that 
the coherent interference effect in the 
time evolution on the grid of anti-crossings under the pulsed electric 
field plays decisive role for the occurrence of such transitional 
behavior.  

The transitional behavior observed here brings us a new powerful method to 
selectively field-ionize the low $\ell$ states from the many close-lying 
states, thus opening a new way to apply the highly excited Rydberg 
states to fundamental physics research.  One of the example of such 
applications is to search for dark matter axions with a Rydberg-atom 
cavity detector~\cite{carrack}. In this kind of search experiment, 
it is essential to do the experiment in a continuous way so as to keep the 
detection efficiency as high as possible. It is thus   
inevitable to use cw lasers to excite the Rydberg atoms 
continuously which is the main reason for developing the present 
experimental setup.                    

The authors would like to thank Akira Masaike for his continuous 
encouragement throughout this work. This research was partly 
supported by a Grant-in-Aid for Specially Promoted Research 
(No.09102010) by the Ministry of Education, Science, Sports, 
and Culture, Japan.


\begin{references}
\bibitem{eff}
Taking into account the velocity distribution of atoms and the detection 
efficiency in the pulsed field ionization, the overall  
detection efficiency is estimated to be 75 \%.
\bibitem{com1}
Note that the sfi detector used is of integral 
type in a sense that the 
sfi signals are integrated with increasing sfi field $v_f$ 
and thus the derivative of the spectra with respect to $v_f$ is 
the true counts at each bin which is shown in 
Fig.~\ref{fig:sfi-spec}.
\end{references}
\end{document}